\documentclass[aps,prc,superscriptaddress,showpacs ]{revtex4-1}
 \usepackage{amsmath}  % ,twocolumn,preprint,
 \usepackage{dcolumn}   % needed for some tables
 \usepackage{makeidx}
 \usepackage{subfig}
 \usepackage{float}
 \usepackage{graphicx,color}
 \usepackage{amsfonts}
 \usepackage[ansinew]{inputenc}
 \usepackage[usenames,dvipsnames]{pstricks}
 \usepackage{epsfig}
 \usepackage{pst-grad}
 \usepackage{pst-plot}
 \usepackage{mathrsfs}

 \def\vect#1{\mbox{\boldmath $#1$}}

               \newcommand{\be}{${^{8}{\rm Be}}$}
             \newcommand{\cc}{${^{12}{\rm C}}$}
              \makeindex

\begin{document}
 \title{Breathing-like excited state of the Hoyle state in \cc}
 \date{\today}
 \author{Bo Zhou}
 \email{bo@nucl.sci.hokudai.ac.jp.}
 \affiliation{Office of International Affairs, Hokkaido University, Sapporo 060-0815, Japan}
 \affiliation{Faculty of Science, Hokkaido University, Sapporo 060-0810, Japan}
 \author{Akihiro Tohsaki}
 \affiliation{Research Center for Nuclear Physics (RCNP), Osaka University, Osaka 567-0047, Japan}
 \author{Hisashi Horiuchi}
 \affiliation {Research Center for Nuclear Physics (RCNP), Osaka University, Osaka  567-0047, Japan}
 \affiliation {International Institute for Advanced Studies, Kizugawa 619-0225,  Japan}	
\author{Zhongzhou Ren}
\affiliation{Department of Physics, Nanjing University, Nanjing 210093, China}
\affiliation{Center of Theoretical Nuclear Physics, National Laboratory of Heavy-Ion Accelerator, Lanzhou 730000, China}
\affiliation{Key Laboratory of Theoretical Physics, Institute of Theoretical Physics, Chinese Academy of Sciences, Beijing 100190, China}
 
\begin{abstract}
The existence of the $0_3^+$ and $0_4^+$ states around 10 MeV excitation energy in \cc\  is confirmed by a fully microscopic 3$\alpha$ cluster model.
Firstly, a GCM (generator coordinate method) calculation is performed by superposing optimized 2$\alpha$+$\alpha$ THSR (Tohsaki-Horiuchi-Schuck-R\"{o}pke)
wave functions with the radius-constraint method. The obtained two excited $0^+$ states above the Hoyle state  are consistent with the recently observed states by experiment.
Secondly, a variational calculation using the single 2$\alpha$+$\alpha$ THSR wave function orthogonalized to the ground and Hoyle states
is made and it also supports the existence of the $0_3^+$ state obtained by the GCM calculation. The analysis of the obtained $0_3^+$ state is made
by studying its 2$\alpha$-$\alpha$ reduced width amplitude, its 2$\alpha$  correlation function, and the large monopole matrix element between
this state and the Hoyle state, which shows that this $0_3^+$ state is a breathing-like excited state of the Hoyle state. This character of the
$0_3^+$ state is very different from the $0_4^+$ state which seems to have a bent-arm 3$\alpha$ structure.
\end{abstract}
 \pacs{21.60.Gx}
\maketitle

As one of the universal phenomena in nature, resonance states widely appear in a large variety 
of fields from particle physics to the condensed matter physics~\cite{moiseyev_non-hermitian_2011}. 
Systems with electrons, hadrons or atoms display various and rich resonances states in different ways, 
which often leads to a new state finding and deepen our understanding for the many-body dynamics. In nuclear physics, due to the complex and special nucleon-nucleon interaction, resonance states are highly common and important in almost all the nuclear 
systems~\cite{ren_new_2004,ni_theoretical_2013,xu_alpha-decay_2016,aoyama_complex_2006}.  As one of most important nuclei in nuclear cluster physics, \cc\ has been studied for a long time  especially for the famous Hoyle state~\cite{freer_hoyle_2014}, which is a narrow 3$\alpha$ resonance state 
and plays a key role in the synthesis of carbon in the universe. In the past decade, it is surprised to find that there 
are quite impressive discoveries and new understanding in this old subject, e.g., many new cluster 
states above the 3$\alpha$ threshold energy were found from experiments like the new 
$0_3^+$, $0_4^+$~\cite{itoh_candidate_2011}, $2_2^+$~\cite{zimmerman_unambiguous_2013}, and 
$4_2^+$~\cite{freer_evidence_2011} states. These observed broad cluster resonance states are expected 
to provide us new clue for understanding the Hoyle state. Actually, as we see in this paper, the $0_3^+$ 
resonance state is intimately related to the Hoyle state. 
 
About 40 years ago, the GCM calculation with the 3$\alpha$ Brink wave function by Uegaki et al.~\cite{uegaki_structure_1979} which reproduced the ground and Hoyle states gave the $0_3^+$ state at $E_x$=11.7 MeV and assigned it to the observed $0_3^+$  state at $E_x$=10.5 MeV . 
Later calculations including the AMD (Antisymmetrized Molecular Dynamics)~\cite{kanada-enyo_structure_2007}  and FMD (Fermionic Molecular Dynamics) ~\cite{chernykh_structure_2007} also gave the $0_3^+$ state around $E_x$=10 MeV.  All the $0_3^+$ states by these calculations have a characteristic feature that they contain non-small component of  \be($2^+$)+$\alpha$(D-wave) configuration. 
In AMD and FMD, this feature has been referred to as the bent-arm structure of 3$\alpha$. However, about 10 years ago Kurokawa and Kat$\bar{\rm o}$ 
reported that the 3$\alpha$ OCM (Orthogonality Condition Model) calculation combined with 
the CSM (Complex Scaling Method) gives another $0^+$ state~\cite{kurokawa_new_2005} around $E_x$=10 MeV in addition 
to the  $0^+$  state with the bent-arm-like structure of 3$\alpha$.  This new  $0^+$  state has 2$\alpha$-$\alpha$
reduced width amplitude whose node number is larger than that of the Hoyle state. The 
existence of two $0^+$  states around $E_x$=10 MeV was soon later supported by Itoh et al.~\cite{itoh_candidate_2011}
experimentally who showed that the broad $0_3^+$ state at $E_x$=10.5 MeV is divided into two 
$0^+$  states, namely $0_3^+$ and $0_4^+$ atates at 9.04 MeV and 10.56 MeV with the widths of 
1.45 MeV and 1.42 MeV, respectively. Itoh et al. reported that the $0_3^+$ state decays 
dominantly through the \be($0^+$)+$\alpha$(S-wave)  channel while the $0_4^+$ state decays through the 
\be($2^+$)+$\alpha$(D-wave)  channel. Thus, the $0_4^+$ state corresponds to the $0_3^+$ state by Uegaki, AMD 
and FMD.  The existence of $0_3^+$ and $0_4^+$ states around 10 MeV was 
reported by another 3$\alpha$ OCM calculation combined with CSM a few years ago~\cite{ohtsubo_complex-scaling_2013} and also 
by a microscopic 3$\alpha$ model calculation last year~\cite{funaki_hoyle_2015}.   Microscopic 3$\alpha$ model calculation 
is especially important for examining the existence of $0_3^+$ and $0_4^+$ states around 10 MeV 
because both AMD and FMD calculations have reported only the existence of $0_4^+$ state.  
It is therefore highly desirable to perform another microscopic 3$\alpha$ model calculation in 
order to confirm the existence of $0_3^+$ and $0_4^+$ states around 10 MeV and also to clarify 
the character of the $0_3^+$ state which is far more unknown than that of the $0_4^+$ state.  

Quite recently, based on the concept of nonlocalized clustering~\cite{zhou_nonlocalized_2013,zhou_nonlocalized_2014}, we proposed an extend THSR wave function which gave a good description of the compact ground states in \cc\ \cite{zhou_container_2014}. In this work, we aim to confirm and investigate the excited  $0_3^+$ and $0_4^+$ states of \cc\ using the extended THSR wave function as basis wave functions in the GCM calculation.  
The extended THSR wave function of Ref.~\cite{zhou_container_2014},  which includes the 2$\alpha$ correlation in 3$\alpha$ cluster structure is written as, 
\begin{eqnarray}
\label{thsr1}
 \Phi(\vect{\beta}_1,\vect{\beta}_2) &=& \int d^3 R_1 d^3 R_2
\exp[-\sum_{i=1}^2(\frac{R_{ix}^2 }{\beta_{ix}^2}+\frac{R_{iy}^2 }{\beta_{iy}^2}+\frac{R_{iz}^2 }{\beta_{iz}^2} )]
\Phi^B(\vect{R}_1,\vect{R}_2) \\
&& \propto \phi_G  {\cal A} \{ \exp[-\sum_{i=1}^2( \frac{\xi_{ix}^2}{B_{ix}^2} +\frac{\xi_{iy}^2}{B_{iy}^2}+
\frac{\xi_{iz}^2}{B_{iz}^2}) ] \phi(\alpha_1)\phi(\alpha_2)\phi(\alpha_3)   \},
\end{eqnarray}
Here, $B_{1k}^2=b^2+\beta_{1k}^2$, $B_{2k}^2=\frac{3}{4} b^2+\beta_{2k}^2$, and $\vect{\beta}_i \equiv(\beta_{ix},\beta_{iy},\beta_{iz})$.  $\vect{\xi}_1=\vect{X}_2-\vect{X}_1$, $\vect{\xi}_2=\vect{X}_3-(\vect{X}_1+\vect{X}_2)/2$. $\Phi^B(\vect{R}_1,\vect{R}_2)$ is the Brink wave function of \cc. $\vect{R}_1$ and $\vect{R}_2$ are the corresponding inter-cluster distance generator coordinates. $\phi_G$ is the center-of-mass wave function of \cc, which can be expressed as, $ \exp(-6X_G^2/b^2)$.   
In practical calculations, we assume the axial symmetry for the 2$\alpha$+$\alpha$ system, namely, $\vect{\beta}_i \equiv(\beta_{ix}=\beta_{iy},\beta_{iz})$ ($i$=1, 2). As for the effective nucleon-nucleon interaction, we adopt the Volkov No.2 force (modified version) with Majorana parameter M=0.59 and  $b$=1.35 fm, which were used by Kamimura et al. for 3$\alpha$ RGM calculation \cite{kamimura_transition_1981}.

As we know, to describe the broad resonance cluster states in \cc, the GCM bound-state approximation is no longer available due to the contamination of the continuum states above the threshold energy.  To remove the contamination, we used the radius-constraint method \cite{funaki_new_2006,funaki_cluster_2015} combined with the GCM. We  diagonalize the squared radius operator and  obtain the corresponding  eigenstates and eigenvalues. Since the larger squared radius eigenvalues indicate the continuum states are involved, we adopt the radius eigenfunctions  whose eigenvalues are smaller than the cutoff parameter $R_{\rm cut}$  in the GCM calculations. This kind of treatment is very similar to the  shell model calculations for  resonance states where nucleon orbits are confined within some radial region.

In GCM calculations, a very large basis is necessary for covering various cluster model spaces for the excited $0^+$ states of \cc. However, considering the numerical errors from GCM combined with radius-constraint method, it is not suitable to superpose directly a huge number of, e.g., more than 1000, THSR wave functions. In this situation, we propose a way for selecting more effective wave functions as the basis. The steps for this one-by-one GCM+RCM (radius-constraint method) are as follows, 
  
(1) We choose a large number of projected normalized $0^+$ THSR wave functions  
\{$\hat{\Phi}^{0^+}_1, \hat{\Phi}^{0^+}_2, \cdots , \hat{\Phi}^{0^+}_{2592} $\} 
as our prepared basis, which correspond 2592 different sets of mesh points 
for $ (\vect{\beta}_1,\vect{\beta}_2) $. The matrix elements of norm, squared radius operator, and Hamilton
are calculated and prepared  for the following calculations. Since the direct diagonalization of Hamilton using this huge prepared basis is very difficult, we want to pick small number of effective wave functions 
one by one from the prepared basis for obtaining converged binding energies 
and wave functions for the ground state and excited $0^+$ states of \cc.

(2) At the beginning, we focus on the ground state of \cc\ and we begin with the first 
effective wave function among the prepared basis. Firstly, we calculate the binding energies 
of single wave functions in the prepared basis.  As for calculations by the single wave function 
$\hat{\Phi}^{0^+}_i$ in GCM+RCM, it simply means if  
$ \sqrt{ \langle \hat{\Phi}^{0^+}_i |(\vect{r}-\vect{r}_{\rm cm})^2/12|\hat{\Phi}^{0^+}_i \rangle}
> R_{\rm cut}$, the wave function $\hat{\Phi}^{0^+}_i$ will be abandoned, otherwise we  
retain this wave function and calculate its binding energy.  Secondly, if some wave function 
like $\hat{\Phi}^{0^+}_{233}$ can give the deepest binding energy for the ground state among the 
prepared basis, then $\hat{\Phi}^{0^+}_{233}$ will be our first selected wave function. 
It should be noted, to obtain the converged value of the ground state, the pure and traditional 
GCM is enough and RCM is not necessary since the bound state is almost independent of 
the parameter $R_{\rm cut}$ in RCM. 

(3) Next, we need to choose the second effective wave function among the prepared basis 
for the ground state of \cc. Assume the first selected wave function is $\hat{\Phi}^{0^+}_{233}$ , 
we then make the diagonalization of Hamiliton for all the superposed two wave functions, 
$ \{ \hat{\Phi}^{0^+}_{233}+\hat{\Phi}^{0^+}_1 \}, \{ \hat{\Phi}^{0^+}_{233}+\hat{\Phi}^{0^+}_2 \}, 
\cdots,\{\hat{\Phi}^{0^+}_{233}+\hat{\Phi}^{0^+}_{2592}\}$ 
using GCM+RCM.  For each ${\hat{\Phi}^{0^+}_{233}+\hat{\Phi}^{0^+}_i} ( i \ne 233)$ we diagonalize 
the squared radius operator and we retain only the eigenfunctions whose eigenvalues are smaller than 
$R_{\rm cut}^2$. If the $\{ \hat{\Phi}^{0^+}_{233}+\hat{\Phi}^{0^+}_{737} \}$ 
group can give the deepest energy for the ground state, then we can choose 
$\hat{\Phi}^{0^+}_{737} $ as our second selected wave function. 
One by one, we can obtain dozens of very effective wave functions (e.g., 50)  
for the ground state and the corresponding eigenvalue has been very well converged. 
Here we emphasize again, as for the selected $n_B$ $0^+$ THSR wave functions from 
the prepared basis in the GCM+RCM calculations, the adopted radius eigenfunctions should 
have smaller ($\leq R_{\rm cut}$) eigenvalues while each of these radius eigenfunctions is 
a linear combination of the selected $n_B$ $0^+$ THSR wave functions. In addition, in the 
selection process, the wave functions bringing fluctuation and large numerical errors for the 
excited $0^+$ states will also be abandoned. 

(4) After selecting 50 effective wave functions for the ground state, in the same way, 
we continue to choose more effective wave functions for the $0_2^+$ , $0_3^+$ , and $0_4^+$ 
states in \cc\ in turn. Namely we select additional effective wave functions so that we get 
deeper binding energies for the $0_2^+$, $0_3^+$ , and $0_4^+$ states. 
Finally, after selecting lots of wave functions with the fixed parameter $R_{\rm cut}$, e.g., the maximum number is around 70 for $R_{\rm cut}$=6 fm, we cannot select any wave functions from the prepared basis for meeting our requirements, then the selection process is completed. 

 \begin{figure}[!h]
 	\centering
 	\includegraphics[scale=0.88]{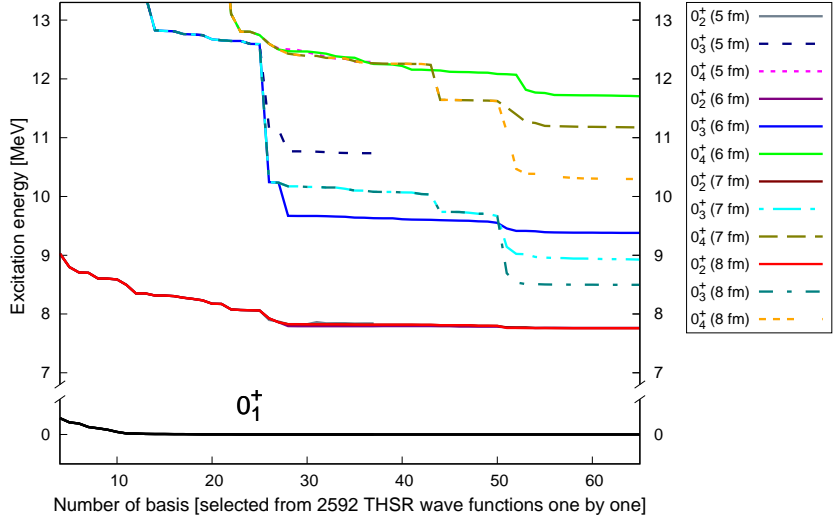}
 	\caption{ \label{fig1} GCM-THSR results for the ground and three excited $0^+$ states of \cc\ using different values of the cut-off parameter $R_{\rm cut}$ in the radius-constraint method.  The values of the cut-off parameter $R_{\rm cut}$ are shown in $0_{k}^+(R_{\rm cut})$ in the insert.  The excitation energies are relative to the obtained GCM converged energy for the  ground state -89.65 MeV \cite{zhou_container_2014}.}
 \end{figure}
 
One-by-one method is a very effective and general approach for selecting the good basis in the 
GCM calculation, especially for some kinds of resonance states with large model spaces.
Figure \ref{fig1} shows the GCM-THSR results for the first four $0^+$ states of \cc\ using 
different values of the radius cut-off parameter $R_{\rm cut}$ in the radius-constraint method. 
The basis wave functions are constructed from 2592 
THSR wave functions (2592 mesh points for ($\beta_{1x},\beta_{1z},\beta_{2x},\beta_{2z}$)). 
It is known that the ground state of \cc\ is a compact bound cluster state and the 
Hoyle state around the 3$\alpha$ threshold energy has a very narrow width of only 8.5 eV, 
which can be seen as a weakly-bound state.  In Fig.~\ref{fig1}, it can be seen that the ground 
state and the Hoyle state in GCM calculations are almost independent of the 
$R_{\rm cut}$ parameter. The energies of the two states reach their converged 
values already at the small number of basis states.  We need to notice that by using the  constructed basis, dozens of superposed wave functions rather than hundreds of them can give a very exact converged solution compared with the traditional GCM calculations.   

As for the broad excited $0^+$ states, the choice of the $R_{\rm cut}$ parameter should be 
careful. The smaller $R_{\rm cut}$ ($\leq$ 5 fm) can lead to the miss of some important 
model spaces while too large $R_{\rm cut}$ ($\geq$ 9 fm ) will bring obvious contamination 
from the continuum states.  The obtained GCM energies of the $0_3^+$  and $0_4^+$ states for $R_{\rm cut}$ = 6 fm 
are seen to be almost  constant against the increase of the number of basis states $n_B$ in the region of $n_B> $ 30 
for the $0_3^+$  state and $n_B> $40 for the $0_4^+$ state.  The constancy of the GCM energy 
against the increase of $n_B$  is a little worse for the $0_4^+$ state than for the $0_3^+$  state, 
but still the constancy of the $0_4^+$ state energy is within the range of about 0.5 MeV. 
The GCM energies of these $0^+$ states for larger $R_{\rm cut}$= 7 fm and 8 fm change their 
values against the increase of $n_B$ although the amount of change is not so large. 
These behaviors of the $0_3^+$  and $0_4^+$ energies for $R_{\rm cut}$= 7 fm and 8 fm mean that 
the GCM wave functions for $R_{\rm cut}$= 7 fm and 8 fm contain the contamination of 
continuum state components.  Thus we conclude that the GCM results for $R_{\rm cut}$= 6 fm 
shown in Fig.~\ref{fig1} give the converged results for the energies and wave functions 
of the $0_3^+$  and $0_4^+$ states. The converged energies 9.38 MeV and 11.7 MeV of the calculated  
$0_3^+$ and $0_4^+$ states, respectively, are consistent with the corresponding observed values 
9.04 MeV and 10.56 MeV of the experimental $0_3^+$  and $0_4^+$ states, respectively. 

 \begin{figure}[!h]
 	\centering
 	\includegraphics[scale=0.78]{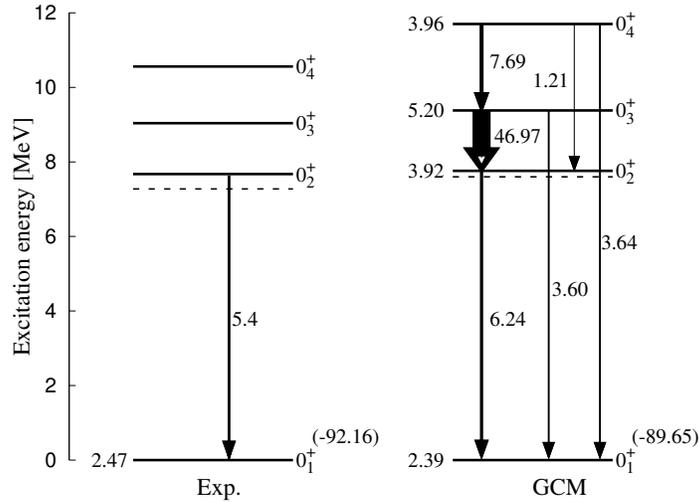}
 	\caption{\label{fig2} The GCM energy levels, r.m.s radii for the mass distributions (left side of the energy levels), and the monopole transition strengths (along the transition lines)  for the ground state and excited $0^+$ states in \cc. The dash lines are corresponding to the threshold energies. It should be noted that the observed radius for the ground state in \cc\ from experiment is charge radius and it is from Ref.~\cite{de_vries_nuclear_1987}. }
 \end{figure}

Next, we show some detailed features of the wave functions of these excited states obtained with ${ R_{\rm cut}}$=6 fm. The GCM energies, r.m.s radii, and the monopole matrix elements are calculated as shown in Fig.\ref{fig2}.
Based on the R-matrix theory~\cite{lane_r-matrix_1958}, the main partial 
$\alpha$-decay widths into ${}^8{\rm Be}(0^+)$ for the  $0_2^+$, $0_3^+$ and $0_4^+$ states are 
calculated to be 7.39 eV, 0.92 MeV, and 0.66 MeV, respectively, which agree with the experimental values 
8.5 eV, 1.45 MeV, and 1.42 MeV for these three excited states. The adopted decay energies measured 
from the decay threshold by which we calculate penetrability factors are taken from experiments.  
The chosen channel radii are 5.5 fm, 10.0 fm, and 4.0 fm, respectively, which give the largest reduced width amplitudes (RWA)  around these points. Thus, the observed data of the $0_3^+$ and $0_4^+$ states
are reproduced by our GCM calculations. It can be seen that the obtained $0_3^+$ state 
has a very large radius, more than 5 fm, which is much larger than the gas-like Hoyle state.  
And the calculated monopole strength between $0_2^+$ and $0_3^+$ of \cc\ is about 47 
${\rm e^2fm^4}$, which is much larger than other monopole transitions. This shows that the broad 
$0_3^+$ state has more dilute density compared with the Hoyle state and we consider that the $0_3^+$ 
state is a kind of breathing excited state of the Hoyle state as we discuss later.

 \begin{figure}[!h]
 	\centering 
 	\includegraphics[scale=0.50,angle=-90]{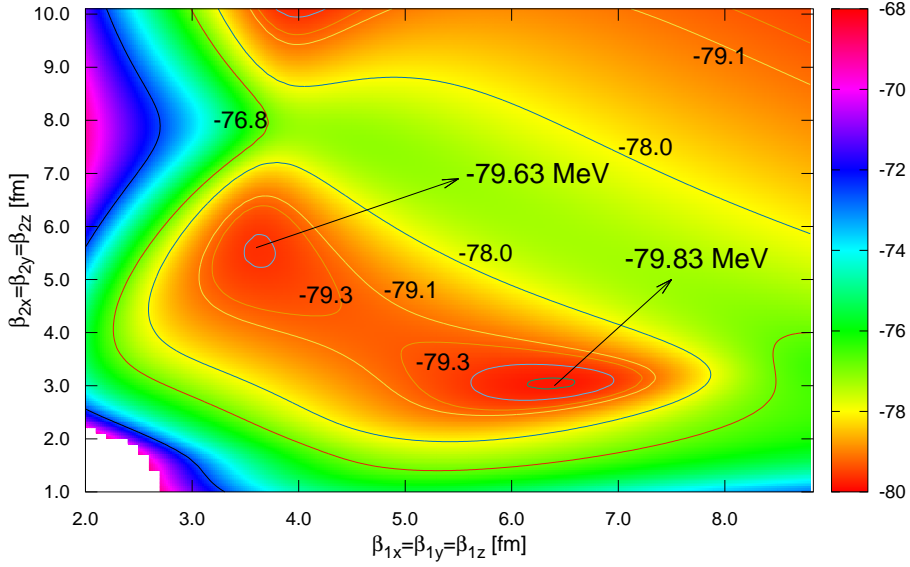}
 	\caption{\label{fig3} The contour plot for the  $0_3^+$ state in the spherical $\vect{\beta}_1$ and $\vect{\beta}_2$ parameter space, which is obtained from the variation calculations of a constructed single THSR wave function  orthogonalized to the ground and Hoyle states of \cc. }
 \end{figure} 

Based on the orthogonality between the $0_1^+$ state and $0_2^+$ state of \cc,  
a single orthogonalized THSR wave function of $0_2^+$ state can be constructed as, 
$\label{02}\hat{\Phi}^{0_2^+}_{2\alpha+\alpha}(\vect{\beta}_1,\vect{\beta}_2)=(1- 
n_1 |\hat{\Phi}^{0_1^+}_{\rm min} \rangle \langle \hat{\Phi}^{0_1^+}_{\rm min}| )
\hat{\Phi}^{0^+} (\vect{\beta}_1,\vect{\beta}_2)$.  
Here, $n_1$ is a normalization factor and 
$\hat{\Phi}^{0_1^+}_{\rm min}(\beta_{1x}=0.1,\beta_{1z}=2.3,\beta_{2x}=2.8,\beta_{2z}=0.1)$ 
is the single optimum THSR wave function obtained by variational calculations, 
which is about 98\% equivalent to the GCM ground wave function \cite{zhou_container_2014}.
Thus, the optimum $0_2^+$ THSR wave function $\hat{\Phi}^{0_2^+}_{\rm min}$ can be obtained 
with the minimum energy 
$E_{\rm min}^{0_2^+}(\beta_{1x}=4.9,\beta_{1z}=2.9,\beta_{2x}=10.7,\beta_{2z}=0.4)$
=-81.79 MeV, which has a 98.3\% squared overlap with the corresponding GCM solution.
In the same way, we can construct a single orthogonalized THSR wave function of $0_3^+$ 
by using the $0_1^+$ and $0_2^+$ wave functions, 
$\hat{\Phi}^{0_1^+}_{\rm min}$ and $\hat{\Phi}^{0_2^+}_{\rm min}$, namely  
$\hat{\Phi}^{0_3^+}_{2\alpha+\alpha}(\vect{\beta}_1,\vect{\beta}_2) =
(1-n_1 |\hat{\Phi}^{0_1^+}_{\rm min}\rangle \langle \hat{\Phi}^{0_1^+}_{\rm min}|
-n_2 |\hat{\Phi}^{0_2^+}_{\rm min}\rangle \langle  \hat{\Phi}^{0_2^+}_{\rm min}|) 
\hat{\Phi}^{0^+} (\vect{\beta}_1,\vect{\beta}_2)$.  
This $\hat{\Phi}^{0_3^+}_{2\alpha+\alpha}(\vect{\beta}_1,\vect{\beta}_2) $ 
wave function provides us another independent and simple way for confirming the
existence of the $0_3^+$ state in \cc. 

Figure \ref{fig3} shows the contour plot for the energy of the $0_3^+$ state as a function of 
spherical $\vect{\beta}_1$ and $\vect{\beta}_2$ calculated by using the wave 
function $\hat{\Phi}^{0_3^+}_{2\alpha+\alpha}(\vect{\beta}_1,\vect{\beta}_2)$.   
The two local minimum points, -79.83 MeV and -79.63 MeV, appear in the contour plot and  
they are connected by a flat valley. The squared overlap between these two states, 
$|\langle \hat{\Phi}^{0_3^+}_{\rm min1}(\beta_1=6.4 , \beta_2=3.0) |
\hat{\Phi}^{0_3^+}_{\rm min2}(\beta_1=3.6 , \beta_2=5.6)\rangle|^2$=0.840, shows 
these two wave functions are not so close compared with the case of the contour plot 
for the Hoyle state \cite{funaki_analysis_2003}.  Above the second local minimum 
point, we have checked that there is a quite large deep region, which belongs to 
the 3$\alpha$ continuum region and there are no local minimum points.  
Furthermore, the first local minimum energy -79.83 MeV is very close  
to the GCM energy -80.27 MeV for the $0_3^+$ state.  Most importantly, 
it is found that the squared overlap between this simple wave function 
$\hat{\Phi}^{0_3^+}_{\rm min1}$ and the GCM $0_3^+$ wave functions, 
$|\langle \hat{\Phi}^{0_3^+}_{\rm min1}(\beta_1=6.4 , \beta_2=3.0) |
\hat{\Phi}_{\rm gcm}^{ 0_3^+}\rangle |^2$, is as high as 0.903. 
If we adopt the deformed THSR wave function, we can find an even better 
wave function and their squared overlap $|\langle \hat{\Phi}_{2\alpha+\alpha}^{0_3^+}
(\beta_{1x}=6.7,\beta_{1z}=4.7,\beta_{2x}=4.1,\beta_{2z}=1.3) 
|\hat{\Phi}_{\rm gcm}^{ 0_3^+}\rangle |^2$=0.944.  
This high squared overlap indicates that the obtained orthogonalized THSR wave function 
$\hat{\Phi}^{0_3^+}_{2\alpha+\alpha}$ for the local minimum energy 
$E_{\rm min}$=-79.8 MeV is just the $0_3^+$ orthogonalized THSR wave function of \cc. 
Thus, we can say that the existence of the $0_3^+$ state is confirmed again using the simple single 
$0_3^+$ THSR wave function orthogonalized to the ground and Hoyle states.

  \begin{figure}[!h]
  	\centering
  	\includegraphics[width=3.6in]{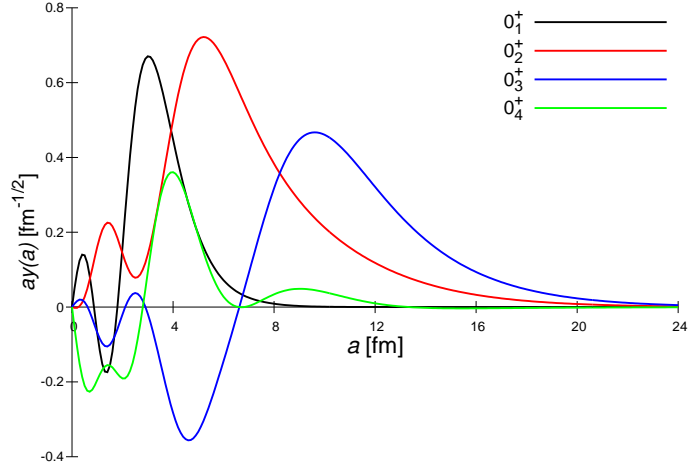}
  	\caption{ 	\label{fig4} The $\alpha$ reduced width amplitudes of the ${0_1^+}, {0_2^+}, {0_3^+},$ and ${0_4^+}$ states for the [\be($0^+) +\alpha]_{0^+}$ channel in \cc.} 
  \end{figure}

Next, using the obtained GCM wave functions, we want to investigate further the $\alpha$+\be\  
correlation of the excited $0^+$ states in \cc. Here, we focus on the domain channel  
[\be($0^+) +\alpha]_{0^+}$ for the ground and excited $0^+$ states in \cc.  
We calculate the $\alpha$ reduced width amplitude (RWA) ${\cal Y}(a)$ defined as,
\begin{equation} {\cal Y}(a)=\sqrt{\frac{12!}{4!8!}} \langle [\hat{\Phi}_{2\alpha}^{0^+},Y_{00}(\vect{\hat \xi}_2)]_{00}
\frac{\delta(\xi_2-a)}{\xi_2^2}\phi(\alpha) |\hat{\Phi}^{0^+_k}_{\rm gcm} \rangle.  
\end{equation}
Here, the normalized projected \be\ THSR wave function is 
$\hat{\Phi}_{2\alpha}^{0^+} \propto P_{00}^{0^+}{\cal A}[e^{-\frac{\xi^2_{1x}}{B_{x}^2}-\frac{\xi^2_{1y}}{B_{y}^2}-\frac{\xi^2_{1z}} {B_{z}^2} }\phi^2(\alpha)]$. 
$B^2_{k}=b^2+\beta^2_k$. In the RWA calculations, $\beta_x=\beta_y=3.0$ fm and $\beta_z=11.1$ fm, 
with which this $2\alpha$ projected THSR wave function gives minimum energy by the use of 
the same interaction parameters of \cc.

Fig. \ref{fig4} shows ${\cal Y}(a)$ for the four $0^+$ states  
($0_1^+ \sim 0_4^+$). It should be noted that, due to the 
optimized basis using one-by-one method in GCM, we got a better and more extended wave functions 
for the excited $0^+$ states in \cc.  We can see that, the $0_3^+$ state has a much larger extension compared 
with the Hoyle state. Since the number of the nodes of the $0_3^+$ state is larger by one than 
that of the Hoyle state, the $0_3^+$ state can be considered as an excited state 
of the Hoyle state at least for 2$\alpha$-$\alpha$ part, which have also been discussed 
in Refs. \cite{uegaki_structure_1979,funaki_hoyle_2015}.  On the other hand, the reduced width amplitude 
of the $0_4^+$ state for the channel \be($0^+) +\alpha $ is much smaller than that of the $0_3^+$ state, 
which implies that the \be($0^+) +\alpha $ component of the $0_4^+$ state is much smaller
than that of the $0_3^+$ state.  This fact is consistent with the bent-arm structure of the $0_4^+$ state.

Another essential problem is, how about the 2$\alpha$ behaviors in these excited states. To study the 2$\alpha$ correlation in the excited $0^+$ states in \cc,  we introduce the following 2$\alpha$ relative wave function of \cc,
 \begin{equation}\label{chi}
 \chi(a)= N_0 \sqrt{\frac{12!}{4!4!4!}}  \langle [e^{-\frac{\xi^2_{2x}}{B_{2x}^2}-\frac{\xi^2_{2y}}{B_{2y}^2}-\frac{\xi^2_{2z}}{B_{2z}^2}  } \phi^3(\alpha)]^{0^+} \frac{\delta (\xi_1-a)}{\xi^2_1}Y_{00}(\vect{\hat{\xi_1}})|\hat{\Phi}^{0^+_k}_{\text {gcm} } \rangle.
 \end{equation}
Here, $N_0$ is the normalization factor $N_0=1/\langle [e^{-\frac{\xi^2_{2x}}{B_{2x}^2}-\frac{\xi^2_{2y}}{B_{2y}^2}-\frac{\xi^2_{2z}}{B_{2z}^2}  } \phi^3(\alpha) ]^{0^+} |   [e^{-\frac{\xi^2_{2x}}{B_{2x}^2}-\frac{\xi^2_{2y}}{B_{2y}^2}-\frac{\xi^2_{2z}}{B_{2z}^2}  } \phi^3(\alpha)]^{0^+} \rangle$. $\chi$(a) is the relative wave function 
between two $\alpha$ clusters inside \cc.  $B_{2k}^2=\frac{3}{4} b^2+\beta_{2k}^2$ and 
their values are chosen as follows.
 
To choose some typical values of the parameter $\vect{\beta}_2$ in Eq.~(\ref{chi}), we firstly search for the largest squared overlaps between the single THSR wave functions and the $0^+$ GCM wave functions. As for the ground state, we have known that the obtained $ \hat{\Phi}^{0_1^+}_{\rm min}(\beta_{1x}=0.1,\beta_{1z}=2.3,\beta_{2x}=2.8,\beta_{2z}=0.1)$ wave function by variational  calculations almost gave the largest squared overlap, 0.978, with the GCM ground wave function.  The obtained largest squared overlaps for the excited $0^+$ states are, $|\langle \hat{\Phi}^{0^+}(\beta_{1x}=9.3,\beta_{1z}=4.6 ,\beta_{2x}=7.2,\beta_{2z}=0.1) |\hat{\Phi}_{\rm  gcm}^{ 0_2^+}\rangle |_{\rm max}^2$=0.837 ; 
$|\langle \hat{\Phi}^{0^+}(\beta_{1x}=9.3,\beta_{1z}=9.2 ,\beta_{2x}=13.8,\beta_{2z}=13.7) |\hat{\Phi}_{\rm  gcm}^{ 0_3^+}\rangle |_{\rm max}^2$=0.290;
$|\langle \hat{\Phi}^{0^+}(\beta_{1x}=0.7,\beta_{1z}=9.2 ,\beta_{2x}=0.7,\beta_{2z}=7.7) |\hat{\Phi}_{\rm  gcm}^{ 0_4^+}\rangle |_{\rm max}^2$=0.446. These obtained largest single THSR wave function components  show that there are possibly very different intrinsic shapes in these excited states in \cc.  For example, the largest component wave function in  $0_3^+$ GCM wave function, $\hat{\Phi}^{0^+}(\beta_{1x}=9.3,\beta_{1z}=9.2,\beta_{2x}=13.8,\beta_{2z}=13.7)$, has a very large and nearly spherical size parameters $\vect{\beta}_1$ and $\vect{\beta}_2$, which reflects the large-radius character of the $0_3^+$ state. And the largest single wave function component in $0_4^+$ GCM wave function has a very obvious deformed prolate intrinsic shape, which indicates the possible rigid bent-arm structures of $0_3^+$ state obtained from AMD and FMD.   

 \begin{figure}[!h]
 	\centering 
 	\subfloat[ ($\beta_{2x}$, $\beta_{2z}$)=(2.8 fm, 0.1 fm)]{%	
 		\includegraphics[width=.45\textwidth]{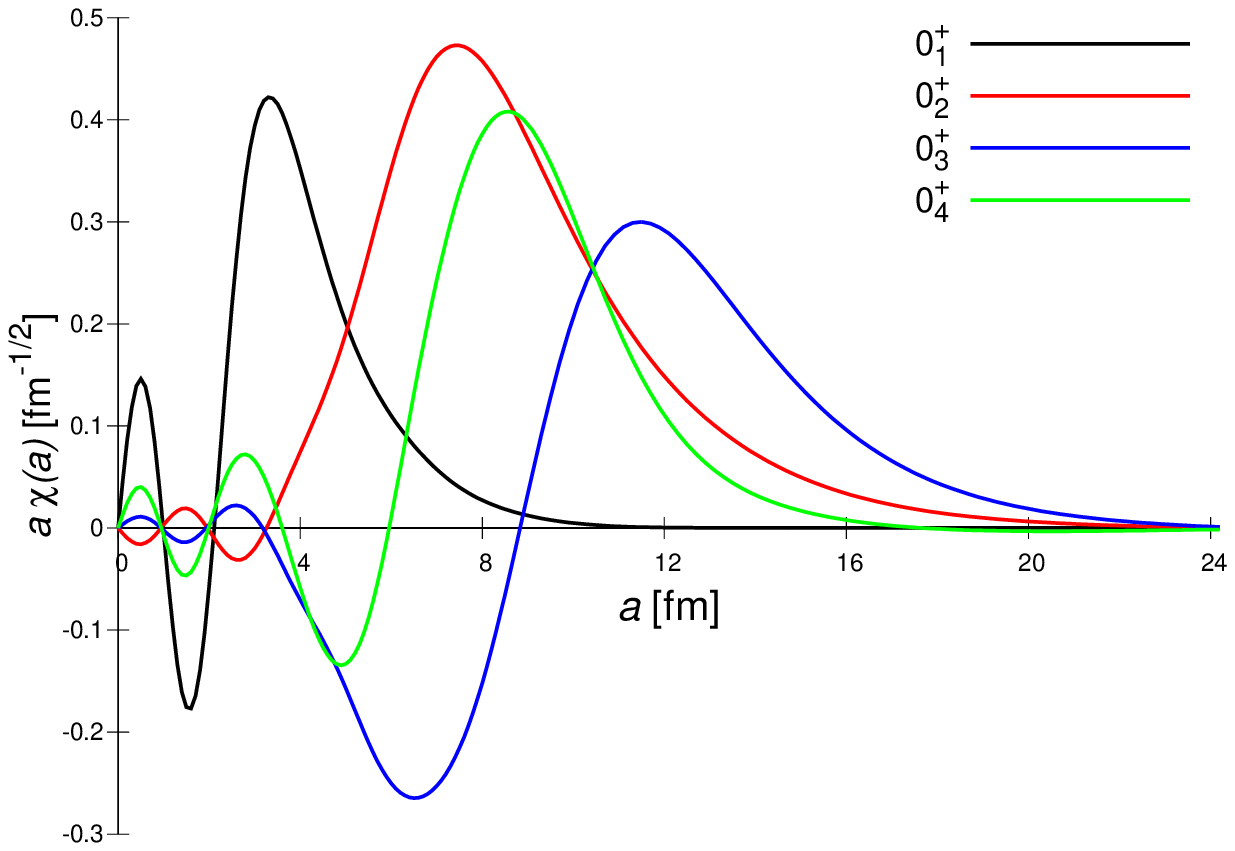}}\hfill
 	\subfloat[ ($\beta_{2x}$, $\beta_{2z}$)=(7.2 fm, 0.1 fm)]{%
 		\includegraphics[width=.45\textwidth]{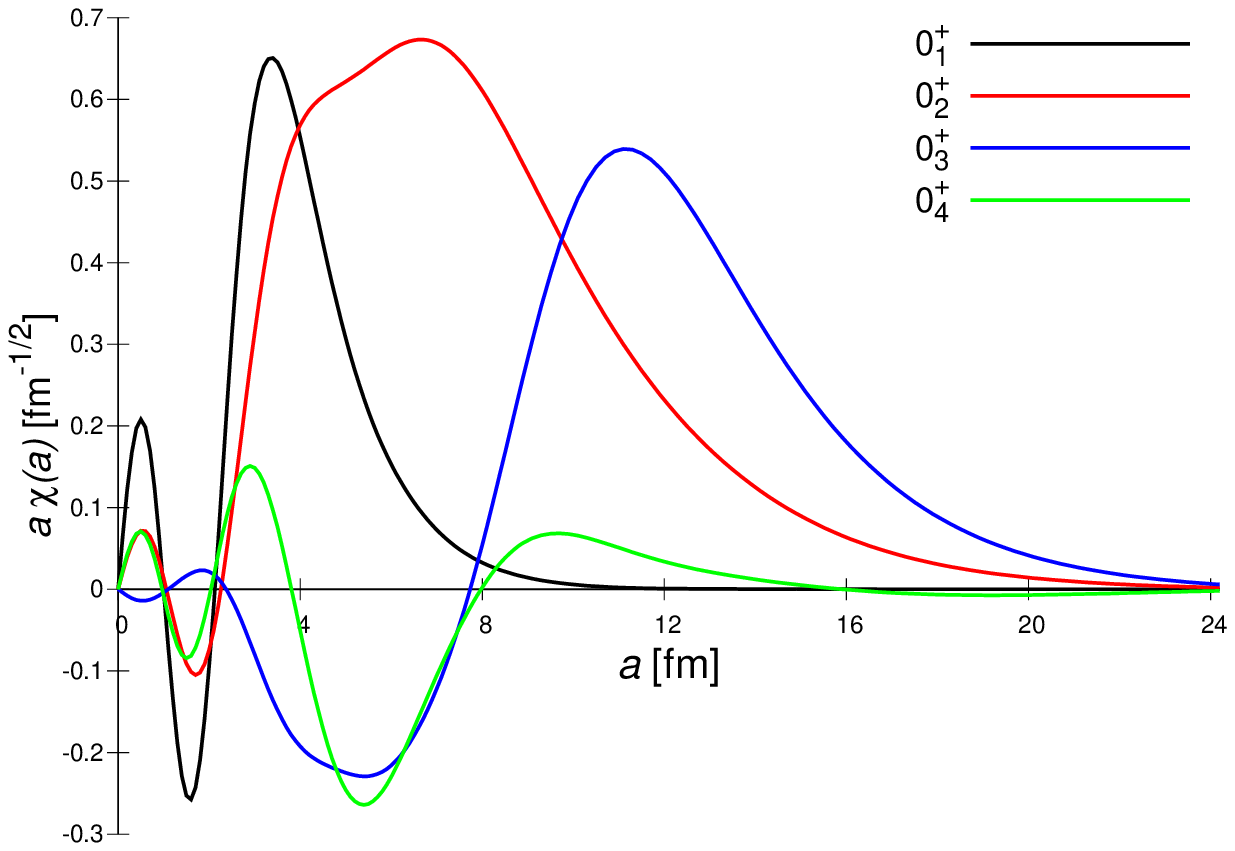}}\hfill
 	\subfloat[ ($\beta_{2x}$, $\beta_{2z}$)=(13.8 fm, 13.7 fm)]{%
 		\includegraphics[width=.45\textwidth]{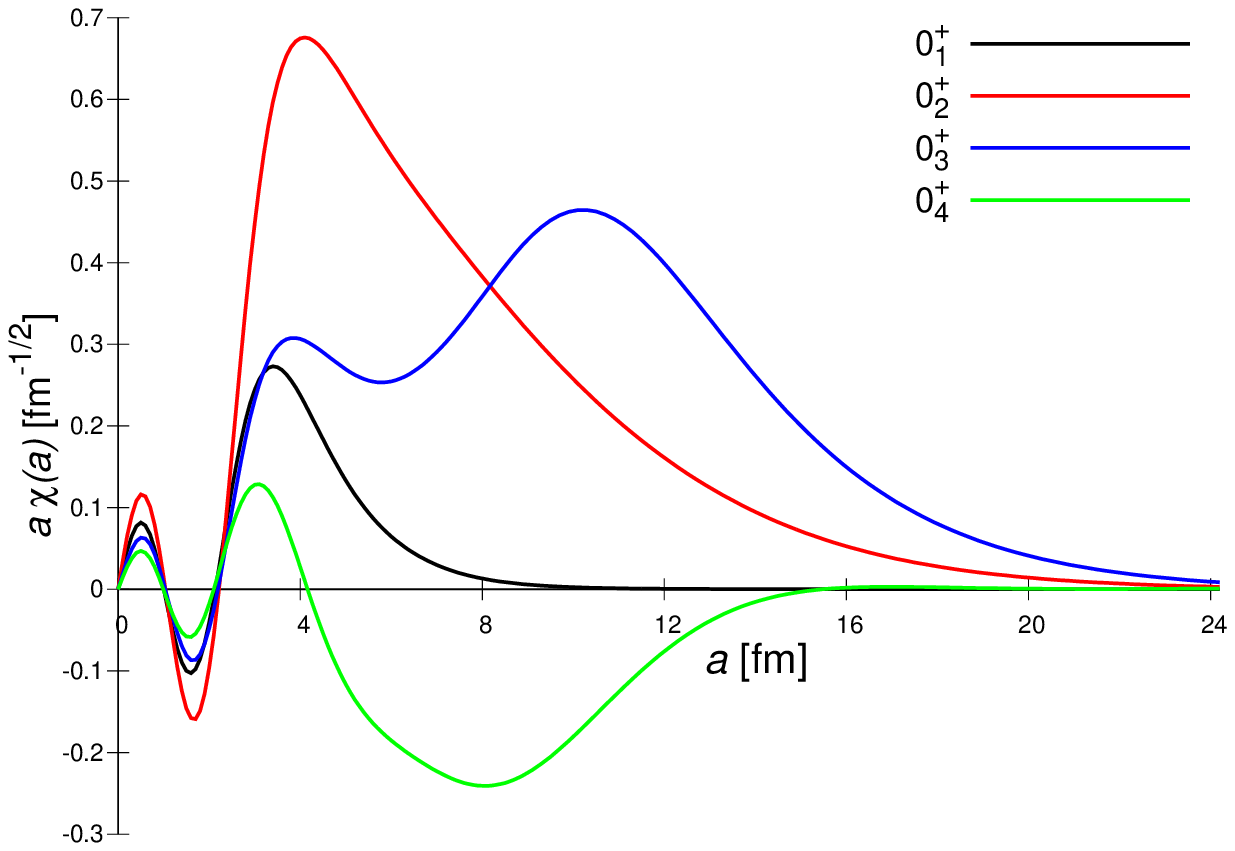}}\hfill
 	\subfloat[ ($\beta_{2x}$, $\beta_{2z}$)=(0.7 fm, 7.7 fm)]{%
 		\includegraphics[width=.45\textwidth]{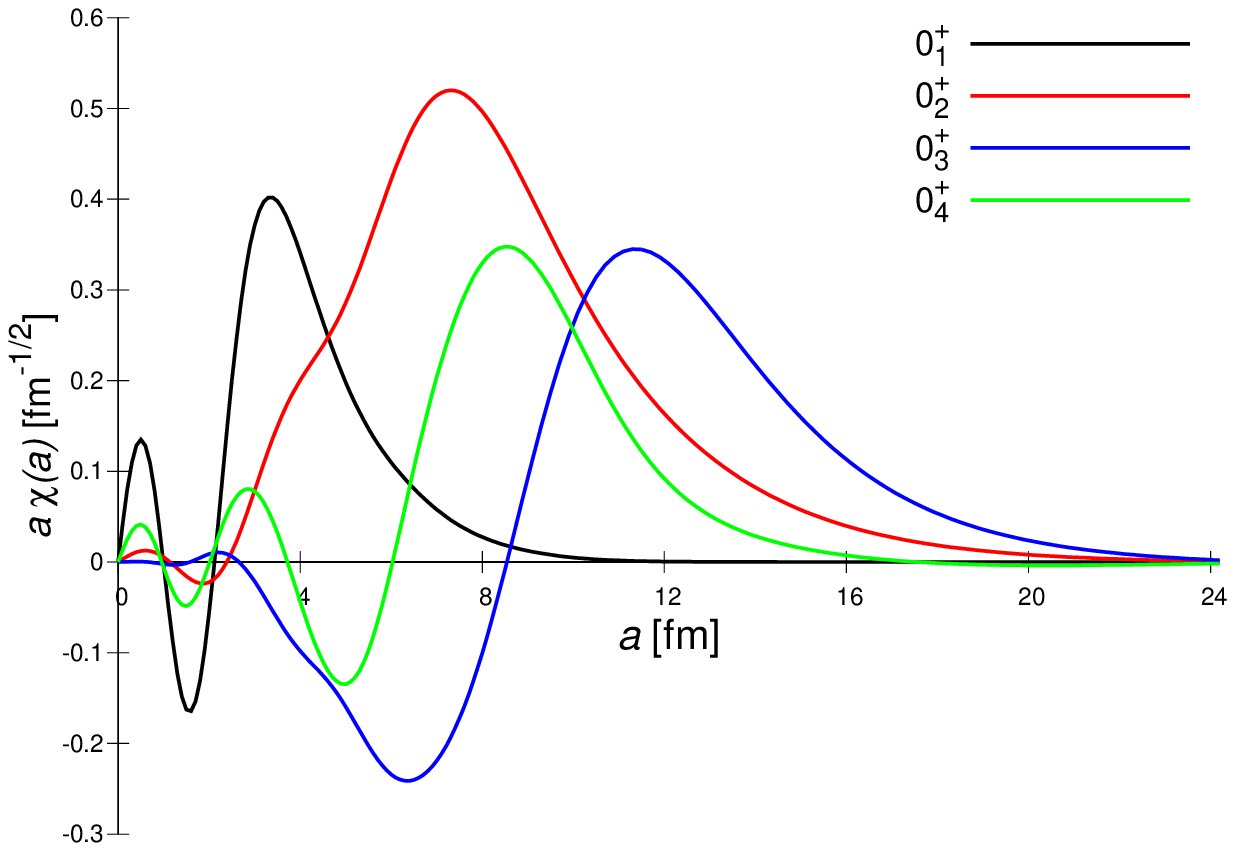}}\\ 
 	\caption{The calculated 2$\alpha$ correlation wave functions of the ${0_1^+}, {0_2^+}, {0_3^+},$ and ${0_4^+}$ states using four sets of $\vect{\beta}_2$ parameters in \cc. }
 	\label{fig5}
 \end{figure}

 To study the 2$\alpha$ correlations of the four $0^+$ states in \cc\ in different situations,  four sets of values of the parameter $\vect{\beta}_2$  in Eq.~(\ref{chi})  are adopted from the above obtained single THSR wave functions. Figure \ref{fig5} shows 2$\alpha$ correlation functions of the ${0_1^+}, {0_2^+}, {0_3^+},$ and ${0_4^+}$ states in \cc\ using different obtained  values of $\vect{\beta}_2$ parameters. It is the first time that the 2$\alpha$ behaviors are calculated in these $0^+$ states in \cc.  
Due to Pauli principle between 2$\alpha$ clusters, in the internal region, the 2$\alpha$ correlation functions have two nodes and they are located at almost the same positions, namely about 1 fm and 2 fm, even for the different $0^+$ states. In the outside region, the 2$\alpha$ correlations in these states display some 
complicated behaviors and how to understand this kind of correlation is the subject of a forthcoming 
paper.  Here, we only want to emphasize that, in the 2$\alpha$ correlation function, for the 
$0_3^+$ state, it has much more extended tail part and also has one more node than 
the Hoyle state in some sense.  It should be noted that, as for Fig. \ref{fig5}(c), the $0_3^+$ state still can be considered 
to have some "node" in outside region of 4 fm $\leq $ a $\leq $ 8 fm, which has the similar situation with 
the Hoyle state in Fig.~\ref{fig4} in the region of 2 fm $\leq $ a $\leq $ 4 fm.
This shows that, compared with the Hoyle state, the ${0_3^+}$ state  is not only excited from the 2$\alpha$-$\alpha$ part but also from 2$\alpha$ correlation part. 

Now, we clarify further the underlying physical meaning of the number of nodes of 2$\alpha$-$\alpha$  and $\alpha$-$\alpha$ relative wave functions for the ${0_3^+}$ state in \cc. 
As we know, the operator which generates the breathing excitation
is just the squared radius operator $O_B$ as follows, 
 \begin{equation}
O_B =\displaystyle{ \sum_{i=1}^{12} }(\vect{r}_i - \vect{r}_{\rm cm})^2.
\end{equation} 
This $O_B$ is nothing but the operator of monopole transition and it also can be rewritten as,
 \begin{equation}
O_B =  \displaystyle{\sum_{k=1}^3 \sum_{i \in \alpha_k} }(\vect{r}_i -\vect{X}_k)^2
+ 2 \xi_1^2 + \frac{8}{3}\xi_2^2,
\end{equation} 
where $\vect{X}_k$ is the center-of-mass coordinate of the $k$-th $\alpha$ cluster.  
The breathing excitation by $\vect{\xi}_2$ coordinate increases the number of nodes of the relative wave function between $2\alpha$ and $\alpha$, while the  breathing
excitation by $\vect{\xi}_1$ coordinate increases the number of nodes of the relative wave function between $\alpha$ and $\alpha$. 
Therefore the breathing excitation is caused by both $\vect{\xi}_1$ and  $\vect{\xi}_2$ coordinates. 
The $2\alpha$-$\alpha$ relative wave function, namely RWA of \cc\ has been discussed for a long time, including the recent work done by Funaki~\cite{funaki_hoyle_2015}. 
However, for regarding the  $0_3^+$  state as a breathing-like excited state of the Hoyle state, we have to study also the $\alpha$-$\alpha$ relative wave function.
In our present paper we investigated, for the first time, the $\alpha$-$\alpha$ relative wave function. 
When we investigate the number of nodes of relative wave functions of $\vect{\xi}_2$ and $\vect{\xi}_1$, we should be careful about the following point.  For example, when we study
the number of nodes of the relative wave function of $\vect{\xi}_2$, the relative wave function of $\vect{\xi}_1$ should be kept non-excited.  The RWA which is the relative wave function of $\vect{\xi}_2$ is calculated by using the ground-state wave function of \be\ for integrating out with respect to $\vect{\xi}_1$. Similarly in calculating the $\alpha$-$\alpha$ relative wave function in Eq.~(\ref{chi}), we used a non-excited relative wave function of $\vect{\xi}_2$, namely simple Gaussian function of $\vect{\xi}_2.$ 
The calculated results for  2$\alpha$-$\alpha$ and $\alpha$-$\alpha$ wave functions in Fig.~\ref{fig4} and Fig.~\ref{fig5} both show that the obtained  $0_3^+$ state can be considered to have one more node than the Hoyle state.  This means that the ${0_3^+}$ state  is not only excited from the 2$\alpha$-$\alpha$ part but also from 2$\alpha$ correlation part. Considering the very large monopole transition from $0_3^+$ state to Hoyle state, therefore, we think this confirmed $0_3^+$ state is a breathing-like mode of the Hoyle state.

In summary, the existence of the $0_3^+$ and $0_4^+$ states in \cc\ is confirmed by using 
an improved THSR-GCM with radius-constraint method. And the existence of the $0_3^+$ 
state is also well supported by variational calculations using the single $0_3^+$ THSR 
wave function orthogonalized to the ground and Hoyle states. Moreover, we found that the $0_3^+$ state has a very large radius and there is a very large monopole transition from this state to Hoyle state. And by showing the RWAs and 2$\alpha$ correlation functions, we found that the $0_3^+$ state is excited from both 2$\alpha$-$\alpha$  part  and  2$\alpha$ correlation part of the Hoyle state. We concluded that  the $0_3^+$ state is a breathing-like excited state of the Hoyle state.

\begin{acknowledgments}
The authors would like to thank Prof. Gerd R\"{o}pke, Prof. Peter Schuck, Prof. Taiichi Yamada, Prof.  Yasuro Funaki, and Prof. Chang Xu for helpful discussions. B.Z. is grateful to the fruitful discussions with Prof. Masaaki Kimura and other members in nuclear theory group in Hokkaido University. Numerical computation in this work was carried out at the 
Yukawa Institute Computer Facility. This work is supported by the National Natural Science Foundation of China (Grants No. 11535004, No. 11375086, No. 11120101005) and also by JSPS KAKENHI Grant Number 16K05351.
\end{acknowledgments}
 %\bibliographystyle{apsrev4-1}
 %\bibliography{myref}
 
 %
 
\end{document}